\documentclass[a4paper]{jpconf}
\usepackage{graphicx}
\usepackage{url}

\begin{document}
\title{Binding in light nuclei: Statistical NN uncertainties vs
  Computational accuracy\footnote{Presented by RNP at Workshop for
    young scientists with research interests focused on physics at
    FAIR 14-19 February 2016 Garmisch-Partenkirchen (Germany)}}

\author{\underline{R. Navarro P\'erez}$^1$, A. Nogga,$^2$ 
J. E. Amaro$^3$ and E. Ruiz Arriola$^3$}

\address{$^1$ Nuclear and Chemical Science Division, Lawrence Livermore
  National Laboratory Livermore, California 94551, USA}

\address{$^2$ Forschungszentrum J\"ulich, Institut f\"ur Kernphysik (Theorie),
Institute for Advanced Simulation, J\"ulich Center
for Hadron Physics and JARA - High Performance Computing, D-52425 J¨ulich, Germany.}

\address{$^3$ Departamento de F\'{\i}sica At\'omica, Molecular y Nuclear
  and  Instituto Carlos I de F{\'\i}sica Te\'orica y Computacional,
  Universidad de Granada  E-18071 Granada, Spain}

\ead{navarroperez1@llnl.gov,a.nogga@fz-juelich.de,
amaro@ugr.es,earriola@ugr.es}

\begin{abstract}
We analyse the impact of the statistical uncertainties of the the
nucleon-nucleon interaction, based on the Granada-2013 np-pp database,
on the binding energies of the triton and the alpha particle using a
bootstrap method, by solving the Faddeev equations for $^3$H and the
Yakubovsky equations for $^{4}$He respectively. We check that in
practice about 30 samples prove enough for a reliable error
estimate. An extrapolation of the well fulfilled Tjon-line correlation
predicts the experimental binding of the alpha particle within
uncertainties.
\end{abstract}


Nuclear structure {\it ab initio} calculations are notoriously
difficult and computationally demanding and have thus so far been
limited to light nuclei, although recently, these calculations have
been extended to more complex systems~\cite{Quaglioni:2015via,
  Meissner:2015una, Dytrych:2016vjy}. Besides giving important input
for applications such as astrophysically relevant nuclear reactions,
these calculations are important tests of current nuclear
interactions.  To this aim, not only the result itself is important
but also the uncertainty~(see e.g. the special
issue~\cite{ireland2015enhancing}.)  From a theoretical point of view
and the inferred predictive power uncertainties can be grouped into
three main categories
\begin{itemize}
\item The input information: the basic nucleon-nucleon (NN)
  interaction should describe a relevant piece of the NN scattering
  data and the simplest two-body bound state: the deuteron.  We will
  call this the statistical uncertainty for reasons to be justified
  below.

 \item The solution method: the way the multinucleon problem is solved
   once the NN interaction is represented. This requires some
   sufficiently high precision which makes computations costly. We
   will call these the numerical uncertainty.
  
 \item The representation problem: the way the input NN data are
   represented theoretically. Normally potentials are used, but the
   form of the potential in the short range region, below $2-3{\rm
     fm}$, is generally not universal, and they are often tailored to
   make the solution of the many body problem as simple as possible.
   We will call these the systematic uncertainty~\footnote{This
     includes in particular any theoretically based expansion of the
     interaction rooted or inspired by QCD such as chiral perturbation
     theory or large $N_c$ expansions were some renormalization scheme
     dependence is unavoidable.}.
\end{itemize}

\begin{figure*}
\centering
\includegraphics[width=\linewidth,height=0.7\linewidth]{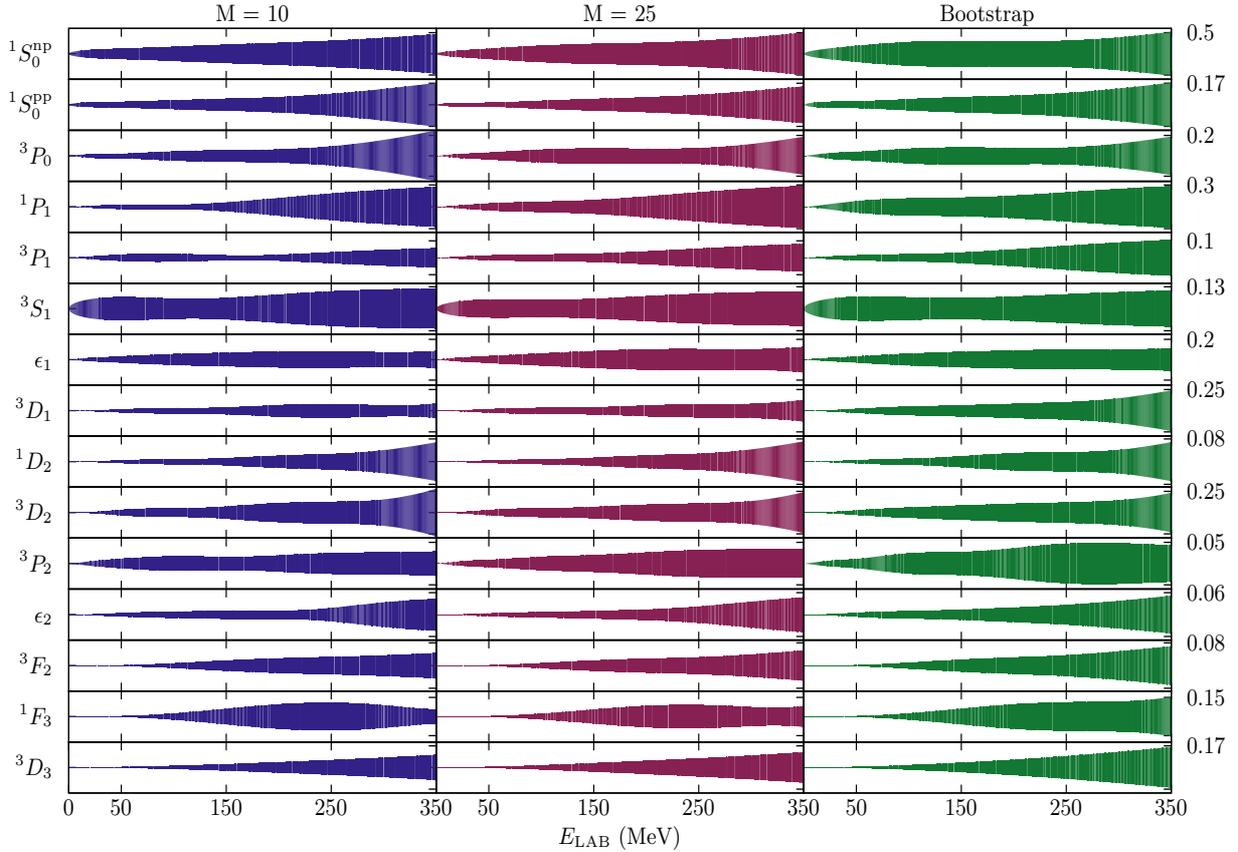}
\caption{(Color online) Phaseshift statistical error bands (in
  degrees) for the $\delta$-shell potential~\cite{Perez:2013jpa}. The
  error bands on the first two columns where obtained using a
  MonteCarlo family of $\delta$-shell potentials where potential
  parameters are random numbers following the multivariate normal
  distribution determined by the original fit covariance matrix. The
  columns use a sample size of $M=10$ (left column) and $M = 25$
  (middle column). The right column is the error bar obtained from the
  Bootstrap to experimental data presented in~\cite{Perez:2014jsa}
  with $M = 1000$. All phase shifts are np unless otherwise indicated.}
\label{fig:PSerrorsDS}       
\end{figure*}

Assuming that these sources of error are independent of each other,  
we expect the total uncertainty to be given, as usual, by 
\begin{eqnarray}
\Delta E^2 = \Delta E_{\rm stat}^2 +\Delta E_{\rm num}^2 +\Delta E_{\rm syst}^2 
\end{eqnarray}
Clearly, the total error is dominated by the largest one. So, it makes
sense either to reduce the largest source of uncertainty or to tune
all uncertainties to a similar level. This sets the limit of
predictive power in {\it ab initio} calculations. While numerical
accuracy has been a goal in itself in few-body calculations, the {\it
  physical} accuracy is given by {\it all} possibles sources of
uncertainties.

In this talk, we discuss the relation between the statistical
uncertainties stemming from the finite experimental accuracy of NN
scattering
data~\cite{NavarroPerez:2012vr,NavarroPerez:2012qf,Perez:2012kt,Perez:2016vzj}
and the currently available numerical accuracy with which the few body
problem can be solved. A pioneering work was carried out in
\cite{adam1993error} where the so-called statistical regularization
was used to evaluate the impact of errors on the binding energies of
the $A=3,4$ systems. The analysis was based on the Paris potential
which has $\chi^2/{\rm d.o.f.} \sim 2$.

The recent Granada-2013 $3\sigma$-self consistent database comprises
6713 np and pp scattering data below $E_{\rm LAB}=350 {\rm MeV}$ and
has a $\chi^2/{\rm d.o.f}=1.04$~\cite{GranadaDB, Perez:2013jpa}. The
procedure to propagate uncertainties is based in spirit on the
bootstrap analysis proposed in \cite{Perez:2014jsa} where the 6713 np
and pp scattering data are randomized and multiple ($M=1020$)
$\chi^2$-fits yield a multivariate distribution of fitting
parameters. This provides a sample enabling a random evaluation of any
observable. We monitor the size $M$ of the needed sample by looking
for statistical stability of the output. The result for the errors in
the corresponding phase shifts is compared in
Fig.~\ref{fig:PSerrorsDS} for different Monte Carlo generated sample
sizes following a gaussian multivariate distribution dictated by the
parameter's covariance matrix. As we see $M=25$ already gives a result
rather close to the full bootstrap method.

\begin{figure}
\centering
\includegraphics[width=0.49\linewidth]{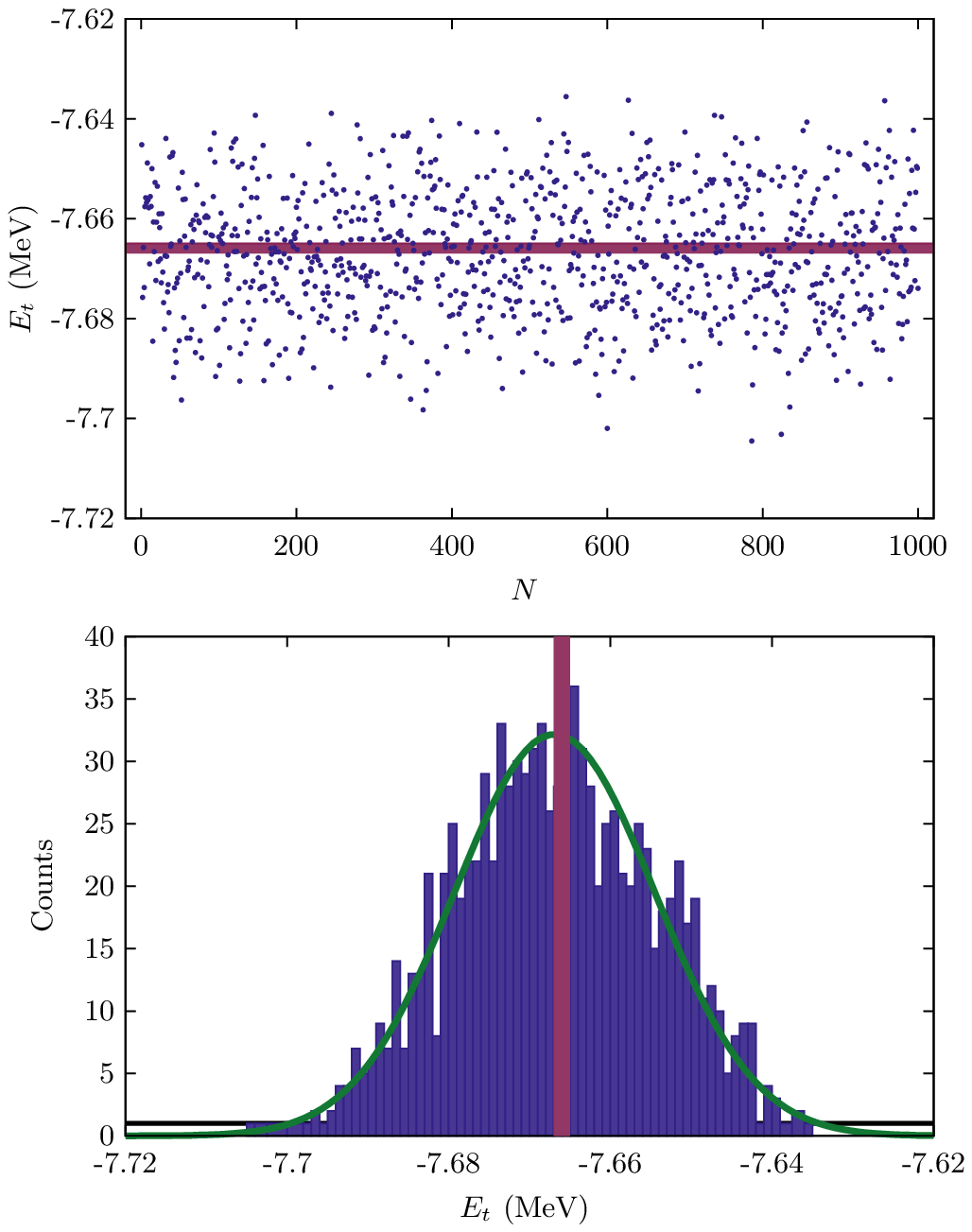}
\includegraphics[width=0.49\linewidth]{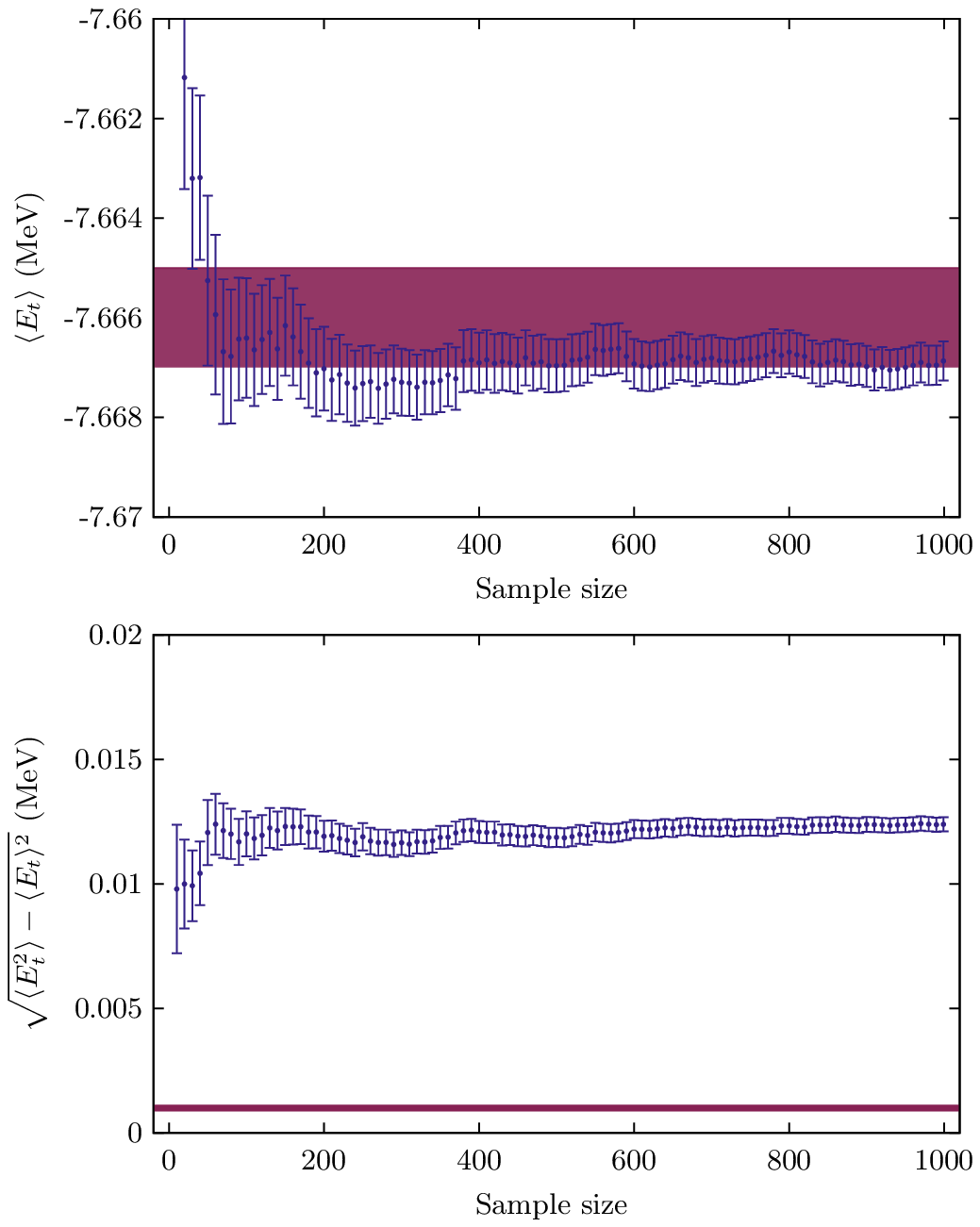}
\caption{(Color online) Distribution (top left) and histogram (bottom left)
  representing the triton binding energy (in MeV) for a sample of
  $1000$ Monte Carlo generated gaussian potential parameters. 
  Finite sample estimates for the population
  mean (top right) and population standard deviation (bottom right) of 
  the triton
  binding energy as a function of the sample size $M$ of the gaussian
  potential parameters are also shown. In all
  panels the red band represents the value obtained with the most
  likely parameters $E_t({\bf p}_0) = 7.666 \pm 0.001 {\rm MeV}$ and
  its width is the numerical error.  We take the bin size equal to
  the numerical precision.}
\label{fig:triton}       
\end{figure}

We have built a simple and smooth gaussian potential which can be
used in most few- and many-body calculational schemes and which
provides an acceptable $\chi^2/{\rm d.o.f.} =
1.06$~\cite{Perez:2014yla}, so it can be considered to be
statistically equivalent to the original delta-shells
potential~\cite{NavarroPerez:2012qf,Perez:2013jpa}.
As we will put forward here, and in agreement with previous findings
using either the hyperspherical harmonics (HSH) method for
$A=3$~\cite{Perez:2014laa} and no-core full configuration shell model
calculations~\cite{Perez:2015bqa}, these estimates already suggests
that the numerical accuracy is close to optimal given the statistical
uncertainty. We will use here the Faddeev equations for the $A=3$
case and the Yakubovsky equations for the $A=4$ situation. As
a first step we will consider only NN forces explicitly and leave out
3N and 4N forces for future developments. The multiple evaluations for
the triton are shown in Fig.~\ref{fig:triton}. As in
\cite{Perez:2014laa} we bin the distribution according to the
numerical accuracy, $\Delta E_t^{\rm num} \sim 1 {\rm
  keV}$~\footnote{One of the advantages of the present method over the
  HSH expansion employed in \cite{Perez:2014laa} is that the
  determination of avoided crossings are difficult to identify
  rigorously; an effect which has to be assisted by visual inspection
  and, if mistaken, has a repulsive effect. This explains the longer
  tails of the triton binding energies distributions. We thank Eduardo
  Garrido for noting this.}.

\begin{figure}[h]
\centering
\includegraphics[width=0.4\linewidth]{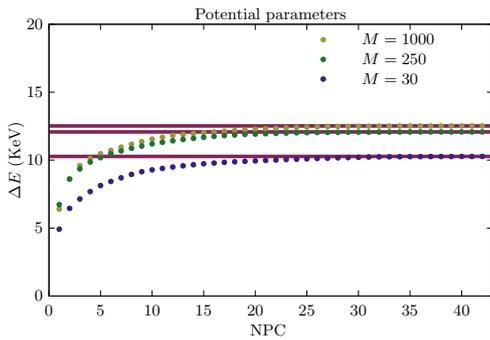}
\begin{minipage}[b]{16pc}
  \caption{(Color online) Number of principal components
    contributions of gaussian parameters for samples of size
    $M=30,250,1000$ (from bottom to top) to the triton binding energy
    uncertainty.
\label{fig:npc-N}}
\end{minipage}
\end{figure}
In a Monte Carlo approach many variations of the parameters produce
irrelevant changes. A principal component analysis looks for eigenvalues
and eigenvectors of the computed observable and provides valuable
information on the most relevant changes of the input parameters but
has seldomly been investigated in nuclear physics (see however
Ref.~\cite{Al-Sayed:2015voa} and references therein). In
Fig.~\ref{fig:npc-N}, we show the results of such an analysis applied
to the coefficients of the gaussian potential of
Ref.~\cite{NavarroPerez:2012qf} implemented in a Monte Carlo fashion.
We found that the number of principal components to obtain most of the
uncertainty in $E_t$ is around $10$. This indicates that regarding
$\Delta E_t^{\rm stat}$ a fit to the NN scattering data base could be
done with less parameters, if the fit were to be designed in
terms of relevant parameters only.

The tiny error band suggests that the discrepancy between our number
$E_t^{\rm th}=-7.6669 \pm 0.0124 {\rm MeV}$ and the $E_t^{\rm exp} =
-8.4820 \pm 0.0001 MeV$ has to be sought in missing three-nucleon
forces (3NFs). It is well known that 3NFs give an important
contribution to nuclear bindings~\cite{Hammer:2012id,
  KalantarNayestanaki:2011wz}. This raises the question of how much of
this statistical uncertainty will be absorbed into variations in the
parameters of the 3NFs. In order to implement some 3N information, we
invoke the empirical linear correlation displayed by the Tjon
line~\cite{Hammer:2012id}~\footnote{See
  \cite{Arriola:2013gya,Arriola:2016fkr} for a Similarity
  Renormalization Group analysis yielding the simple formula $E_\alpha
  = 4 E_t - 3 E_d$.}.

\begin{figure}
\centering
\includegraphics[width=0.49\linewidth]{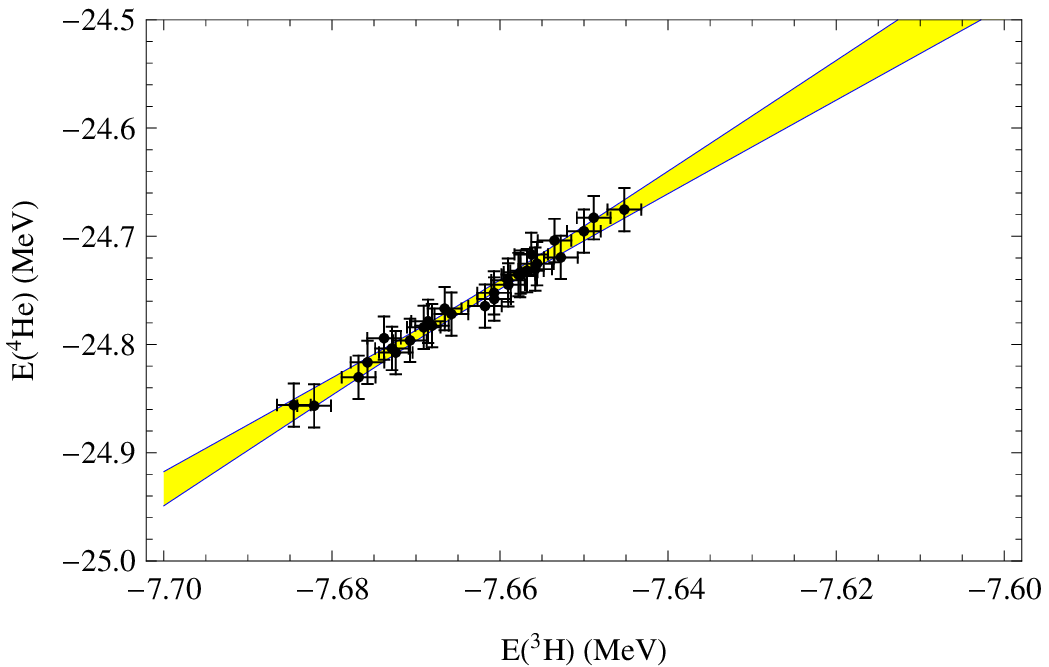}
\includegraphics[width=0.49\linewidth]{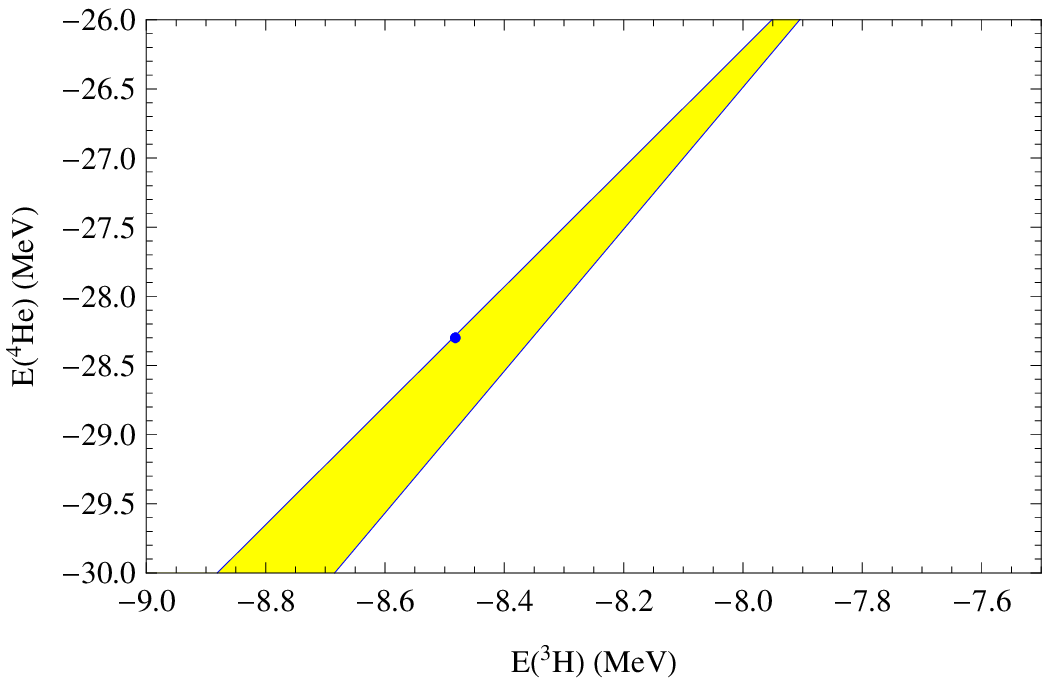}
\caption{(Color online) Tjon type analysis of the $^4$He binding
  energy vs the $^3$H binding energy. We show the fit to the sample of
  $N=30$ Monte Carlo generated binding energies both in a small scale
  (left panel) and extrapolated in a larger scale (right panel)
  compared with the experimental point (blue dot). We take $\Delta E_t^{\rm num}
  =1 \, {\rm keV}$ and $ \Delta E_\alpha^{\rm num} =20 \, {\rm keV}$.}
\label{fig:tjon}       
\end{figure}

In the Monte Carlo method, any choice of parameters ${\bf p}$
determines a value of the triton binding energy. Given the variations
of the triton binding energy, we expect, when determining the $\alpha$
particle binding energy, a Tjon-like linear correlation of the form $
E_{\alpha}( {\bf p} ) = a E_t ( {\bf p)} + b $. The values found are
$a=4.7(1)$ and $b=11.4$.  Thus we expect a Tjon-like correlation would
give $\Delta E_\alpha^{\rm stat} = 4.7(1) \times \Delta E_t^{\rm stat}
= 50(5) {\rm keV}$ which is mainly determined by the channels
involving relative S-waves. In Fig.~\ref{fig:tjon}, we show our
results for the $^3$H and $^4$He binding energy. The yellow band shows
the fit including the uncertainty. The error bars show the numerical
uncertainty. Whereas the variation of the binding energies is rather
large, the linear correlation indicates that most of this variation
will be eventually absorbed into a properly adjusted 3NFs. We take the
band width as an indication for the remaining error induced by the
uncertainty of NN data. As one can see, the band width and the
numerical errors are comparable. Therefore, we deduced that this
uncertainty is comparable to the statistical one. Strong efforts to
increase the numerical accuracy are therefore not desired. For this
analysis, we used a moderate sample of only M=30, the smallness of
which is justified from the analysis of Fig.~\ref{fig:triton}, as far
as uncertainty estimates are concerned. In order to have a tighter
predicted extrapolated band one would need to reduce the numerical
error in $E_\alpha$ in harmony with the Tjon slope $ \Delta
E_\alpha^{\rm num} \sim 4.7 \Delta E_t^{\rm num} \sim 5 {\rm
  keV}$. Further details will be presented in a forthcoming
publication.

\section*{Acknowledgements}
This work is supported by Spanish DGI with Feder funds (grant
FIS2014-59386-P) and Junta de Andaluc{\'{\i}}a (grant FQM225), the
U.S. Department of Energy by Lawrence Livermore National Laboratory
under Contract No. DE-AC52-07NA27344, the U.S.  Department of Energy,
Office of Science, Office of Nuclear Physics under Award
No. DE-SC0008511 (NUCLEI SciDAC Collaboration).  The numerical
calculations have partly been performed on JUQUEEN, JUROPA and JURECA
of the JSC, J\"ulich, Germany.
\section*{References}
\bibliography{references}

\end{document}